%% file: main.tex
\documentclass[manuscript]{acmart}
\AtBeginDocument{%
  \providecommand\BibTeX{{%
    \normalfont B\kern-0.5em{\scshape i\kern-0.25em b}\kern-0.8em\TeX}}}

\setcopyright{acmcopyright}
\copyrightyear{2018}
\acmYear{2018}
\acmDOI{XXXXXXX.XXXXXXX}

\acmConference[Conference acronym 'XX]{Make sure to enter the correct
  conference title from your rights confirmation emai}{June 03--05,
  2018}{Woodstock, NY}
\acmBooktitle{Woodstock '18: ACM Symposium on Neural Gaze Detection,
 June 03--05, 2018, Woodstock, NY} 
\acmPrice{15.00}
\acmISBN{978-1-4503-XXXX-X/18/06}

\usepackage{graphicx}
\usepackage{float}
\usepackage{enumitem}
\usepackage[normalem]{ulem}
\usepackage{hyperref}
\usepackage{tcolorbox}
\usepackage{multirow}
\usepackage{multirow}
\usepackage{fontawesome5}
\usepackage{svg}

\newcommand{\papertitle}{Toward a Scalable Census of Dashboard Designs in the Wild:\\ A Case Study with Tableau Public}

\newenvironment{tight_itemize}{\begin{itemize} \itemsep
-2.1pt}{\end{itemize}}

\definecolor{chocolate}{rgb}{0.48, 0.25, 0.0}

\definecolor{clus0}{HTML}{1f77b4}
\definecolor{clus1}{HTML}{ff7f0e}
\definecolor{clus2}{HTML}{2ca02c}
\definecolor{clus3}{HTML}{d62728}
\definecolor{clus4}{HTML}{9467bd}
\definecolor{clus5}{HTML}{8c564b}
\definecolor{clus6}{HTML}{e377c2}
\definecolor{clus7}{HTML}{7f7f7f}
\definecolor{clus8}{HTML}{bcbd22}
\definecolor{clus9}{HTML}{17becf}
\definecolor{clus10}{HTML}{BC5821}
\definecolor{clus11}{HTML}{D3BFFF}
\definecolor{clus12}{HTML}{4AA0AD}
\definecolor{clus13}{HTML}{F9CC70}
\definecolor{clus14}{HTML}{86D6B4}
\definecolor{clus15}{HTML}{ED6A5A}
\definecolor{clus16}{HTML}{8D0327}
\definecolor{clus17}{HTML}{8B8C64}
\definecolor{clus18}{HTML}{D49A6A}
\definecolor{clus19}{HTML}{306367}
\definecolor{clus20}{HTML}{03396C}
\definecolor{clus21}{HTML}{E85641}
\definecolor{clus22}{HTML}{451e3e}
\definecolor{clus23}{HTML}{1f77b4}
\definecolor{clus24}{HTML}{ff7f0e}
\definecolor{clus25}{HTML}{2ca02c}
\definecolor{clus26}{HTML}{d62728}

\usepackage{adjustbox}

\definecolor{chartcolor}{HTML}{6699CC}
\definecolor{textcolor}{HTML}{FF6633}
\definecolor{filtercolor}{HTML}{669933}
\definecolor{legendcolor}{HTML}{993399}
\definecolor{multimediacolor}{HTML}{996600}

\newtcbox{\blockTypeBox}{nobeforeafter, colback=gray!003, colframe=gray!25, boxrule=0.5pt, arc=1pt, boxsep=0pt,left=2pt,right=2pt,top=1.75pt,bottom=1.5pt,tcbox raise base}
\newcommand{\chart}{\blockTypeBox{{\textcolor{chartcolor}{{\small{\faChartBar}}~chart}}}}
\newcommand{\chartB}{\blockTypeBox{{\textcolor{chartcolor}{{\small{\faChartBar}}~chart}}}}
\newcommand{\charts}{\blockTypeBox{{\textcolor{chartcolor}{{\small{\faChartBar}}~charts}}}}
\newcommand{\chartsB}{\blockTypeBox{{\textcolor{chartcolor}{{\small{\faChartBar}}~charts}}}}

\newcommand{\textTmp}{\blockTypeBox{{\textcolor{textcolor}{\raisebox{-.1em}{\includegraphics[height=0.95em]{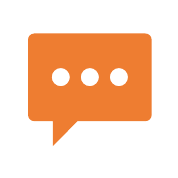}}~text}}}}
\newcommand{\textB}{\blockTypeBox{{\textcolor{textcolor}{\raisebox{-.1em}{\includegraphics[height=0.95em]{figures/icons/text-icon.pdf}}~text}}}}

\newcommand{\filter}{\blockTypeBox{{\textcolor{filtercolor}{{\footnotesize{\faFilter}}~filter}}}}

\newcommand{\filters}{\blockTypeBox{{\textcolor{filtercolor}{{\footnotesize{\faFilter}}~filters}}}}

\newcommand{\legend}{\blockTypeBox{{\textcolor{legendcolor}{{\footnotesize{\faPalette}}~legend}}}}
\newcommand{\legends}{\blockTypeBox{{\textcolor{legendcolor}{{\footnotesize{\faPalette}}~legends}}}}

\newcommand{\multimedia}{\blockTypeBox{{\textcolor{multimediacolor}{{\small{\faImage}}~multimedia}}}}
\newcommand{\multimediaB}{\blockTypeBox{{\textcolor{multimediacolor}{{\small{\faImage}}~multimedia}}}}

\begin{document}

\title{\papertitle}


\author{Joanna Purich}
\authornote{Both authors contributed equally to this research.}
\affiliation{%
  \institution{University of Maryland, College Park}
  \city{Maryland}
  \country{USA}}
\email{banana@umd.edu}

\author{Arjun Srinivasan}
\authornote{Both authors contributed equally to this research.}
\affiliation{%
  \institution{Tableau Research}
  \city{Washington}
  \country{USA}}
\email{arjunsrinivasan@tableau.com}

\author{Michael Correll}
\affiliation{%
  \institution{Tableau Research}
  \city{Washington}
  \country{USA}}
\email{mcorrell@tableau.com}

\author{Leilani Battle}
\affiliation{%
  \institution{Paul G. Allen School of Computer Science \& Engineering}
  \city{Washington}
  \country{USA}}
\email{jpkumquat@consortium.net}

\author{Vidya Setlur}
\affiliation{%
  \institution{Tableau Research}
  \city{Washington}
  \country{USA}}
\email{vsetlur@tableau.com}

\author{Anamaria Crisan}
\affiliation{%
  \institution{Tableau Research}
  \city{Washington}
  \country{USA}}
\email{acrisan@tableau.com}

\renewcommand{\shortauthors}{Trovato and Tobin, et al.}

\include{sections/00-abstract}




\maketitle

\input{sections/01-introduction}
\input{sections/02-related-work}
\input{sections/03-data-structure}
\input{sections/04-methods}
\input{sections/05-results}
\input{sections/05b-interaction-analysis}
\input{sections/06-applications}
\input{sections/07-discussion}
\input{sections/08-conclusion}


\bibliographystyle{ACM-Reference-Format}
\bibliography{main}

\appendix

\end{document}

%% file: sections/00-abstract.tex
\begin{abstract}
Dashboards remain ubiquitous artifacts for presenting or reasoning with data across different domains.
Yet, there has been little work that provides a quantifiable, systematic, and descriptive overview of dashboard designs \emph{at scale}. We propose a schematic representation of dashboard designs as node-link graphs to better understand their spatial and interactive structures. We apply our approach to a dataset of 25,620 dashboards curated from Tableau Public to provide a descriptive overview of the core building blocks of dashboards in the wild and derive common dashboard design patterns. To guide future research, we make our dashboard corpus publicly available and discuss its application toward the development of dashboard design tools.
\end{abstract}

%% file: sections/01-introduction.tex

\section{Introduction}
Dashboards are used broadly to visually explore, monitor, and communicate with data~\cite{wexler2017big,eckerson2010performance,wilbanks2014review,roberts2017give,yigitbasioglu2012review}.
While initial applications of dashboards focused on their use ``to rapidly monitor current conditions''~\cite{few2006information}, recent work has expanded our understanding of how they are used, such as for supporting decision-making, persuading, or learning~\cite{sarikaya_what_2019,bach2022dashboard}. 
However, this newfound versatility also challenges the limits of what current visualization tools and techniques can support. A close examination of dashboards across a variety of real-world use cases and intents is crucial not only for researchers seeking to understand dashboards as a design idiom but also for creating tools to assist in dashboard authoring.

Despite their ubiquity, little work has been done to deeply investigate this wide range of dashboard designs and their behaviors at scale. 
Existing analyses focus on qualitatively coding a hand-picked set of dashboards from different sources across the internet, including news websites, dashboard galleries, and social media~\cite{sarikaya_what_2019,bach2022dashboard}.
While this approach reveals a rich dashboard design space, we observe two key limitations to this approach. First, it may not capture the significance of specific design patterns within the design space, especially vernacular practice among non-expert designers.
As a result, prior analyses may miss key dashboard design practices, or fail to address the most pervasive design challenges that users encounter in their daily work.
Second, manual inspection of dashboards limits the range of these studies to at most a couple of hundred dashboards, a tiny fraction of the dashboards observed in the wild. Similarly, tools and features developed from this narrow perspective might miss key preferences and design patterns among users. In short, while existing dashboard work has created a few hand-picked ``zoos'' of interesting examples (compare to Heer et al.'s~\cite{heer2010zoo} ``visualization zoo'' of ``more exotic (but practically useful) forms of visual data representation''), to our knowledge there is less focus on what a ``census'' of dashboards might be: analysis of both the quotidian and exotic forms of dashboards, at scale and usefully representative of a wide population of designs.

To address these limitations,
we investigate 25,620 dashboards on Tableau Public\footnote{\url{https://public.tableau.com/en-gb/s/}}, a repository contributed to and shared by a wide audience of visualization designers with diverse backgrounds, perspectives, and skill sets. However, exploring and understanding dashboards at such a scale is non-trivial, and involves answering a series of open questions including:

\begin{itemize}
    \item [\textbf{RQ1)}] How can we represent dashboards such that we can compare, contrast, and analyze their design \emph{at scale}?
    \item [\textbf{RQ2)}] Can we leverage this representation to:
    \begin{itemize}[leftmargin=0in]
        \item [\textbf{a)}] identify what are the visual components of dashboards and structural relationships between these components?
        \item [\textbf{b)}] check whether interactivity is common in dashboards and understand how interaction manifests in dashboards?
        \item [\textbf{c)}] detect and characterize high-level dashboard design patterns?
    \end{itemize}
\end{itemize}

To address these research questions, we synthesize a ``census'' of dashboards that enables us to explore the breadth of what is possible with the prevalence of current practices, both effective and ineffective, and ultimately support the creation of better dashboard authoring tools or design guidelines.
Specifically, we formalize dashboards as graphs connecting ``blocks'' of design elements and leverage this formalism to explore the repository of Tableau dashboards.
We analyze both the visual and interactive features of the dashboards to identify prominent design patterns.
Our findings complement prior work by highlighting that dashboards ``in the wild'' are not simply static monitoring devices but can be heterogeneous artifacts composed of a rich set of design elements, including text, filtering widgets, and images.
However, our analyses also reveal design practices unobserved in previous studies, such as a tight coupling of story-driven text elements with visualization elements within dashboards,  the widespread use of canonical visualization types across dashboard designs, and the prevalence of interaction in dashboards.
We discuss potential applications of our dataset and formalization toward designing dashboard tools and understanding authoring practices.
Lastly, distilling the findings from our analyses, we discuss design implications for dashboard authoring systems, including increased consideration for non-visualization elements, supporting richer formatting capabilities, and incorporating onboarding strategies for interactive dashboards.

Our work provides both a \textit{scalable} and \textit{diverse} analysis of dashboards: we can employ quantitative methods that allow analysis at the level of thousands of dashboards, and our corpus contains dashboards designed at multiple levels of expertise from different domains, and constructed for different purposes.

In summary, this research makes the following contributions:
\begin{tight_itemize}
    \item A schematic representation of dashboards as node-link graphs representing the core design elements as well as spatial and interactive relationships between these elements.
    \item Descriptive statistics summarizing visual elements and interactive features of dashboards in the wild generated by applying the proposed schematic representation to the Tableau Public repository.
    \item Cluster analysis of dashboards on Tableau Public highlighting prevalent visual and interactive design patterns along with exemplar dashboards illustrating these patterns.
    \item A publicly available\footnote{A website link will be added to the paper after the reviewing cycle. We currently provide the corpus as part of the \textbf{supplementary material} hosted at: \url{https://osf.io/r5cfk/?view_only=0cca76d27dac402496646c20359c3da6}} dataset of 25,620 Tableau dashboards containing the (anonymized) dashboard contents along with their spatial and interactive relationships.
\end{tight_itemize}



%% file: sections/02-related-work.tex
\section{Related Work}

\subsection{Dashboards as Objects of Study}

Sarikaya et al.~\cite{sarikaya_what_2019} point to an apparent disconnect between the ubiquity of dashboards in visualization \textit{practice} and their lack of consideration in visualization \textit{study}. More recent work has sought to remedy this gap by 1) further clarifying the various forms and goals of dashboard designers and users and 2) codifying or testing design rules or recommendation systems for automating aspects of dashboard design.

Exploring dashboards (and other visualization practices) is often done through an analysis of dashboards in a particular context of use or population of users~\cite{dourish}. For instance, Tory et al.~\cite{tory2021finding} explore dashboard usage among ``data workers,'' i.e., people who do not self-describe as analysts but who nonetheless perform everyday work with data. While valuable, these analyses require access to both the people and visualizations they work with and rely on qualitative and subjective judgments of intent or goal, limiting both the scale and generalizability of results. For instance, Sarikaya et al.~\cite{sarikaya_what_2019}, Bach et al.~\cite{bach2022dashboard} and Al-Maneea et al.~\cite{al-maneea_towards_2019} all explore dashboards and multiple view visualizations with an eye toward their visual structure and topology, but rely on a manual process of coding dashboard features and connections. These manual inspections are valuable and afford inferences about qualitative information that would be difficult to determine automatically. However, we speculate that there are also benefits to what Moretti~\cite{moretti2000} calls ``distant reading'' approaches to corpus analytics, where computational methods afford the identification of large-scale patterns and trends. Distant reading then allows the more principled selection of individual exemplars for more manual inspection and ``close reading.'' See \autoref{sec:related-corpora} for more discussion of these trade-offs. 

The analysis and observation of dashboards are often performed in order to compare these dashboards to existing design guidelines or recommendations from both the academic and practitioner communities~\cite{sarikaya_what_2019,chen2020composition,roberts2007state,wang2000guidelines,qu2017keeping,al-maneea_towards_2019,bach2022dashboard}. For instance, Qu and Hullman~\cite{qu2017keeping} examine how users attend to design inconsistencies between visualizations within the same dashboard and translate their observations into explicit design guidelines for keeping coordinated visualizations consistent. Kristiansen et al.~\cite{kristiansen_semantic_2022} extend these consistency constraints by allowing users to specify \emph{relations}, i.e., subsets of visualizations within a dashboard, and expose constraint parameters to increase or decrease consistency among a relation. Similarly, Langner et al.~ \cite{langner_multiple_2019} perform an observational study of dashboard use and design in large display environments to inform the design of their coordinated view system. Other dashboard authoring or recommendation systems, especially those that use machine learning to attempt to create meaningful layouts and content~\cite{ma2020ladv,odonovan_learning_2014,Oppermann2021vizcommender,wu_learning_2021}, rely on having access to substantial training corpora of well-designed or useful dashboards.

\subsection{Analyses of Visualization Corpora}
\label{sec:related-corpora}
Analyses of large corpora of visualizations have been performed for a variety of reasons. For instance, to describe the flexibility of a specific tool and the habits of its users~\cite{viegas2007manyeyes,battle2018beagle,morton2014public}, to create and evaluate datasets for training machine learning models~\cite{hu2019vizml,hu2019viznet}, or to simply to enumerate the sheer diversity and structure of a design space~\cite{jena2020uncertainty,schulz2011treevis,tominski2017timeviz}. While our motivations span these categories, we note specific structures in how these corpora are collected and analyzed.

Existing corpora can be divided along three dimensions: manual~\cite{sarikaya_what_2019,al-maneea_towards_2019} versus automated~\cite{hu2019vizml,hu2019viznet,battle2018beagle,chen2021vis30k,isenberg2017vispubdata,lee2018viziometrics,morton2014public,hoque2020searching} data collection, manual~\cite{sarikaya_what_2019} versus automated~\cite{hu2019vizml,hu2019viznet,hoque2020searching,davila2021survey} annotation (or both~\cite{battle2018beagle}), and analyzing visualizations as static~\cite{battle2018beagle,hu2019vizml,hoque2020searching} versus dynamic~\cite{sarikaya_what_2019} (i.e., interactive) objects. Each dimension involves trade-offs in the richness, scope, and quality of analyses supported by the annotated data. For example, automated data extraction allows for thousands of examples to be collected, but often with limited annotation, as it is difficult to write programs (e.g. web crawlers) to handle the heterogeneity exhibited in massive corpora. This issue often leads to a relatively limited set of features available for analysis, based on what extraction and annotation programs can reliably detect \textit{en masse}~\cite{battle2018beagle,hoque2020searching}. Ensuring that a massive corpus does not contain data that are out-of-scope, irrelevant, or otherwise inappropriate for downstream analysis is also difficult and may require explicit validation steps~\cite{hu2019viznet}. In contrast, manual data collection and annotation can lead to richer input data and thus a wider variety of potential analyses~\cite{sarikaya_what_2019}, but sacrifice scale in return since manual data collection and annotation involve significant expenditures of time and effort.

We observe one exception - when a large corpus can be extracted using a consistent format, richer \emph{and} broader analyses become possible. The closest example is the VizML project~\cite{hu2019vizml}; since all of the visualizations analyzed were Plotly visualizations, they could be processed in a consistent format, allowing for in-depth extraction and annotation of one million visualization designs and ultimately an effective input dataset for training deep learning models to make encoding decisions. That being said, prior work focuses primarily on collecting and annotating \emph{individual visualizations}, and richly annotated corpora for dashboards are rare and/or limited in size and detail. We present a unique opportunity to analyze and share thousands of dashboard designs in a consistent format amenable to systematic, quantitative analysis.

\subsection{Graph-Based Analysis of Visualization Designs}
Ease in authoring and analyzing visualizations is often linked to the way that a visualization is \textit{specified}. For instance, declarative specification languages like Vega or Vega-Lite~\cite{satyanarayan2016vega} can afford direct analysis of the design of a visualization (such as the type of encodings used or the interactive components present) through simple lookups. 
When the initial specifications are not readily available, one could use alternative techniques, such as image segmentation~\cite{battle2018beagle,davila2021survey,poco2017,poco2018,prasad2007classifying,savva2011} to derive approximate representations of visualizations. However, the heterogeneity of visualization images---and thus, their approximate representations---limits our ability to analyze them at scale. It is also hard to precisely extract higher-level semantics such as layout and interaction properties from images without rich metadata.
In our work, given that we are interested in the \emph{relationships} that bind discrete elements within a dashboard together, we rely on \emph{graph-based representations} for our analysis. 

A number of works explore graph-based representations of visualization and dashboard designs. For example, visualization recommendation algorithms often represent the visualization design space as a graph, where nodes represent specific encoding or data transformation choices and edges reflect relationships between these design decisions~\cite{zeng_evaluation-focused_2022,moritz_formalizing_2019,crisan_gevitrec_2021}. Dashboard designs can also be represented as a graph to capture relationships between different elements, such as directional relationships between interactions in one element that change the encodings or data transformations in another element~\cite{eichmann2020idebench, purich2022adaptive}. Our work explores potential types of graph structures and the notion of a connection between components in a dashboard.

%% file: sections/03-data-structure.tex
\begin{figure*}[t!]
    \centering
    \includegraphics[width=\textwidth]{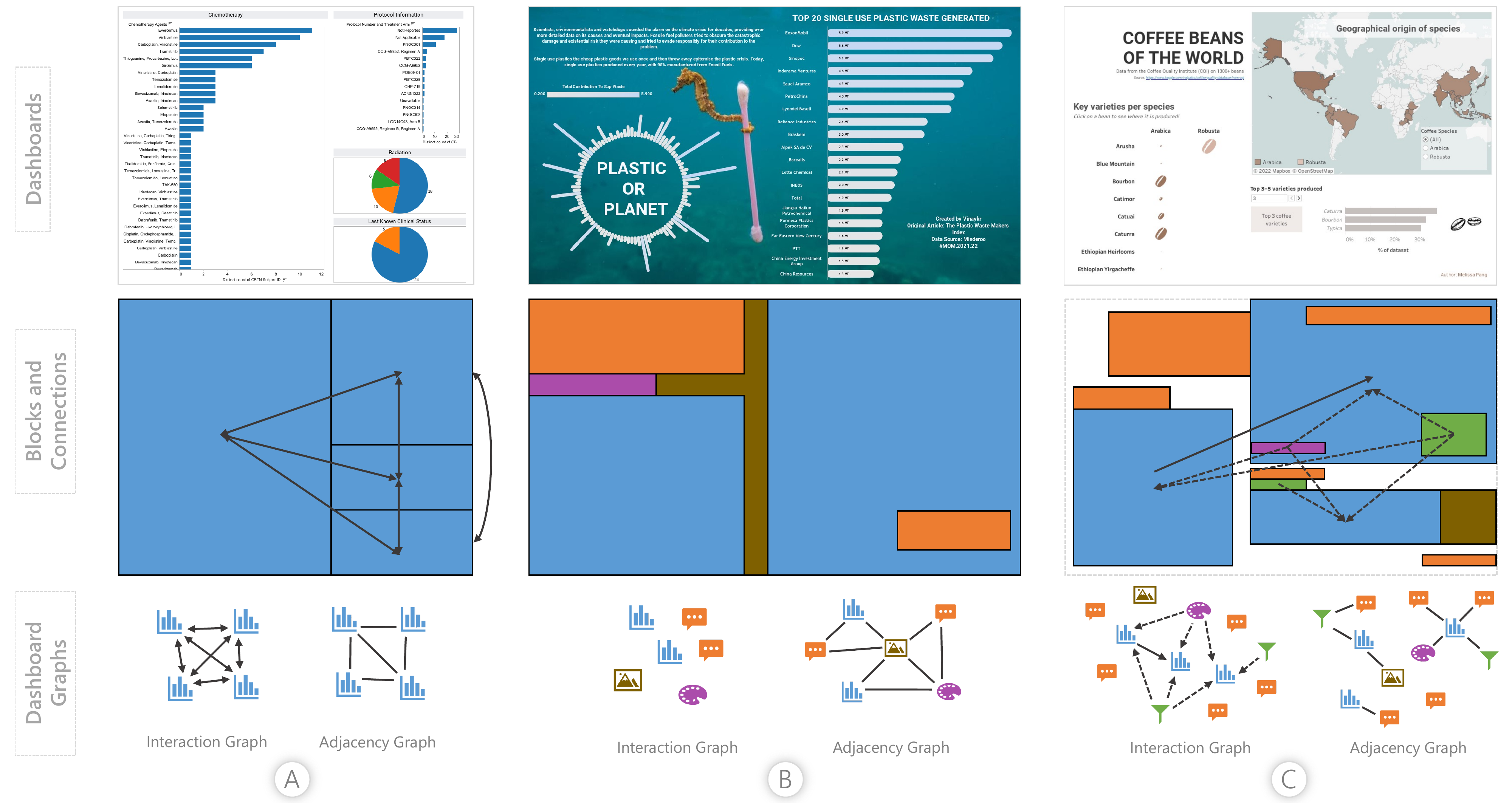}
    \caption{
    Example dashboards are decomposed into block-connection form and reconstructed as interaction and adjacency graphs.
    Blocks are iconified and colored by type (\chart, \legend, \filter, \textB, \multimedia).
    Directed edges show interactions between blocks, while undirected edges indicate that two blocks are adjacent.
    Solid links in the block diagrams and interaction graphs represent {\chart}$\rightarrow${\chart} interactions, and dashed links represent {\filter}$\dashrightarrow${\chart} or {\legend}$\dashrightarrow${\chart} interactions.
    Since (B) has no interactive blocks, its interaction graph contains no edges.
    Likewise, the adjacency graph in (C) is disconnected due to the usage of white space between blocks.
    }
    \label{fig:all-examples}
\end{figure*}

\section{Schematic Representation of Dashboard Design}
\label{sec:dashboard-structure}

To address \textbf{RQ1} (\textit{``How can we represent dashboards such that we can compare, contrast, and analyze their design at scale?''}), we propose a schematic representation of dashboards as two node-link graphs.
In this section, we first make a case for having a general-purpose schematic representation of dashboard designs and then detail our proposed representation.

\subsection{Why a Schematic Representation?}
Dashboards can be specified in myriad ways, including using graphic design tools such as Illustrator~\cite{illustrator2022}, visualization tools like Tableau~\cite{tableau2022}, or programmatically via toolkits like d3.js~\cite{bostock2011d3} or Vega-Lite~\cite{satyanarayan2016vega}.
This flexibility in the authoring process results in a wide spectrum of dashboard designs with varied uses, including data analysis, exploration, interpretation, and presentation.
Figure~\ref{fig:all-examples} highlights a small slice of this diversity in dashboard designs and presents three dashboards that are composed of different visual elements, including but not limited to data visualizations, and with different levels of interactivity.

Figure~\ref{fig:all-examples}A displays a classic multiple coordinated views-style dashboard that is designed for interactive exploration of cancer treatment statistics.
The dashboard comprises four charts (two bar charts and two pie charts).
The bar chart on the left displaying names of chemotherapy agents makes up half of the dashboard, and the other three views displaying protocols and treatments are used in a complementary manner.
All four views in this dashboard are linked through interactions such that clicking on a bar or an arc slice in any of the charts filters the other three charts to show data for the selected category.

Figure~\ref{fig:all-examples}B presents a dashboard with additional elements besides visualizations.
The dashboard discusses single-use plastic and its impact on the ocean and is composed of two visualizations, two text blurbs (with one overlaid onto a visualization), a color legend, and a background image.
Unlike the first dashboard, this dashboard is static and does not support any interactions between its components, and is in service of passive information dissemination rather than active data exploration or real-time monitoring.

Finally, the dashboard in Figure~\ref{fig:all-examples}C displaying data on coffee beans around the world, contains an even broader mix of elements, including visualizations, images, text, legends, and a filtering widget to select the number of categories that can be shown in the bar chart in the bottom-right corner. The dashboard also contains more complex interactions both between visualizations (i.e., cross-filtering) in addition to visualization and other widget elements (i.e., selecting a coffee type in the legend updates the data shown in the charts)

To gain insight into these different dashboards and their underlying design choices, it is useful to describe, compare, and contrast the visualizations, design components such as text and images, and the interactive elements that comprise each of these dashboards.
One way to do so would be to consolidate a summary of their designs and formalize them into a framework or design space description---an approach adopted by prior work~\cite{sarikaya_what_2019,bach2022dashboard}.
However, this approach requires human labor to construct and organize artifacts. The manual and subjective nature of this process makes it challenging to apply it at scale and thus may prevent understanding the frequency of use of different elements and whether some design patterns are more prevalent than others.
Between the low-level specification of individual dashboards and a high-level design space compiled from a qualitative survey of a small set of dashboards, we lack a nuanced and scalable analysis of dashboard designs.
Performing such an analysis at scale, however, requires a consistent representation of dashboard designs that can be used to compare dashboards agnostic of their creation method and tool.

\subsection{Proposed Schema: Blocks and Connections}\label{sec:formalism}
\label{sec:dashboard-structure:formalism}

Our schematic representation treats dashboard designs as a node-link graph comprising a set of \textbf{\textit{blocks}} (nodes) and \textbf{\emph{connections}} (links) between these blocks, where:

\begin{description}
 \item[Blocks] represent the visual elements of a dashboard. Blocks not only include visualizations, but any visual component of the dashboard, including but not limited to text, legends, filter widgets, and multimedia elements such as images or embedded web pages.
 \item[Connections] describe the spatial layout of elements and interaction topology of a dashboard as pairwise relationships between blocks. Specifically, two blocks are connected if 1) they share a common edge or overlap spatially (e.g., the bar chart on the left in Figure~\ref{fig:all-examples}A is adjacent to all other charts, the text block on the bottom-right corner in Figure~\ref{fig:all-examples}B is overlaid on the bar chart), or 2) if the dashboard is configured to be interactive such that interacting with items in one block results in changes in another block (e.g., selecting an item in any chart in Figure~\ref{fig:all-examples}A filters the other three charts).
\end{description}

\begin{figure}[t!]
    \centering
    \includegraphics[width=.6\linewidth]{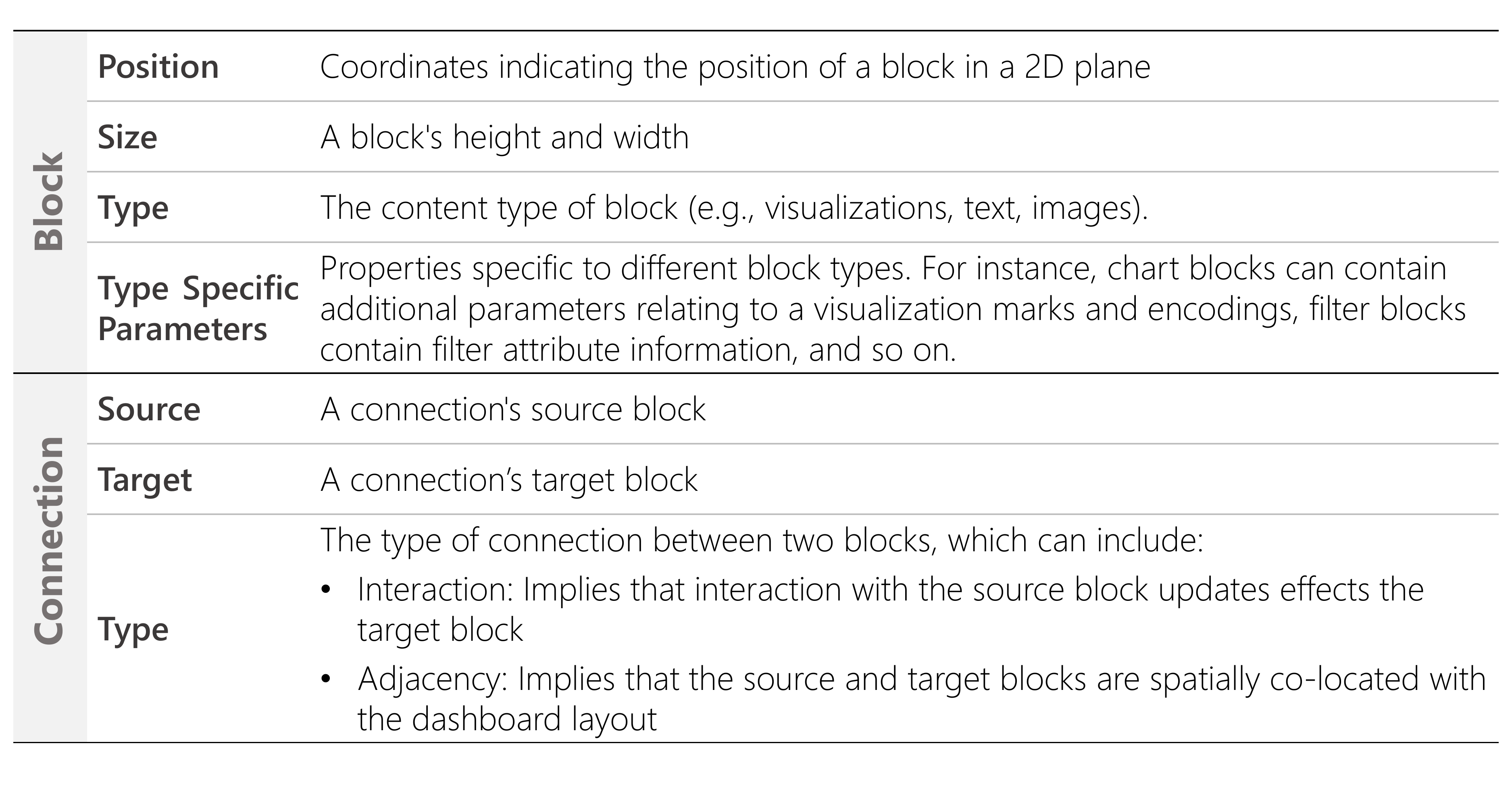}
    \caption{Summary of properties of dashboard blocks and connections between blocks.}
    \vspace{-5mm}
    \label{fig:data_structure}
\end{figure}

Blocks and connections record a wide range of dashboard design \textbf{\emph{properties}}, as shown in~\autoref{fig:data_structure}. Block properties can be split into two groups: \textit{positional} and \textit{descriptive}. Positional properties, including 2D coordinates (\textit{x}, \textit{y}) and size (\textit{w}, \textit{h}), are used to understand the visual layout of blocks.
Depending on the block type, additional descriptive properties are included to further specify the contents of the block.
For instance, for visualization blocks, descriptive properties include the visualization type, the types of data attributes and encodings they employ, and so on.
Similarly, for blocks containing text or filters, the descriptive properties include the actual text string along with any formatting specifications applied to the text or the type of widget (e.g., a dropdown, slider) and data field used as the filter, respectively.

Connections contain properties capturing both the \textit{source} and \textit{target} block for a connection and the \textit{type} of connection that is defined between them.
Just as with blocks, additional type-specific parameters can be utilized to capture supplemental information such as the interaction type (e.g., filter or highlight). 

\subsubsection{Connecting Blocks to Form Graphs}

Together, blocks and connections enable the expression of a wide variety of dashboard designs in a consistent way that can be used for scalable analysis.

For the analysis, we can further model dashboards as 
two types of graphs that reflect both the design and interactive properties of the graph. These multiple graph structures can mutually contribute to holistic analyses such as clustering or pattern detection~\cite{tang2009multigraphcluster}.

The first graph is an \textbf{\emph{adjacency graph}}, an undirected graph that codifies the spatial layout of a dashboard. In this graph, blocks represent nodes and the spatial proximity of blocks establishes edges (connections). The spatial structure of the dashboard enables us to understand not only the types of blocks that dashboard authors use but also whether there exist any consistent design patterns to this structure (\textbf{RQ2a}: \textit{``What are the visual components of dashboards and what are the structural relationships between these components?''}).
These design patterns may be extracted from edges created between two consistent block types or revealed as cliques or neighborhoods in the graph.

The second graph is an \textbf{\emph{interaction graph}}: a directed graph showing how blocks filter or update one another. Nodes in the interaction graph also represent individual blocks, but edges are derived from the interactions between two blocks as opposed to their spatial proximity. By maintaining interactions as a separate graph, the directionality of interaction effects is preserved, enabling the detection of commonly used interactive elements and interaction patterns (\textbf{RQ2b}: \textit{``Is interactivity is common in dashboards? How does interaction manifest in dashboards?''}). However, the two complementary graphs may still be viewed collectively to explore the intersection of layout and interactivity.

\subsubsection{Leveraging Dashboard Graphs for Scalable Design Analysis}

By focusing on the core elements of dashboards and how they relate to one another, our schematic representation not only generalizes to most common dashboard designs, it is also \emph{agnostic} to the tools and languages used to create them. Furthermore, when blocks and their connections can be extracted programmatically from dashboards, our representation enables \emph{scalable} analysis of dashboard designs.


A wide range of dashboard creation properties can be inferred from the resulting graphs.
As an illustration, we apply the proposed schema of blocks and connections, and display the interaction and adjacency graphs for the three dashboards in Figure~\ref{fig:all-examples}.
We notice that decomposing the dashboards into blocks and codifying them into different types, such as \charts{}, \textB{}, and \legends{}, among others, makes it conducive to observing the differences in the visual elements across dashboards (\textbf{RQ2a}).
In this case, for example, the dashboard in Figure~\ref{fig:all-examples}A features only \chartsB, while both Figure~\ref{fig:all-examples}B,~\ref{fig:all-examples}C contain more complex combinations of multiple block types.

The schematic representations also make it easier to understand and compare the spatial (\textbf{RQ2a}) and interactive configuration of the different dashboards (\textbf{RQ2b}).
For instance, the adjacency graph for the dashboard in Figure~\ref{fig:all-examples}C highlights the spatial associations between the different \textB{} and \chartB{} blocks, making it easier to understand which text corresponds to which chart.
Similarly, the interaction graphs not only help in observing the number of interactions but also the directionality and the type of interactions in a dashboard.
For example, the directed links in the interaction graph at the bottom of~\autoref{fig:all-examples}A explicitly highlight the 12 interactions between the four charts in the dashboard.
The absence of links in the interaction graph for~\autoref{fig:all-examples}B indicates that the dashboard is static.
Lastly, the eight links in the interaction graph of~\autoref{fig:all-examples}C highlight that the dashboard is primarily driven by the \legend{} and \filter{} blocks but has one \chart{} (the coffee bean heatmap at the bottom-left corner) that can be used to drive another \chart{} (the map on the top-right corner).

Note that it is straightforward to map the core elements of our representation to their counterparts in existing visualization tools. For example, one can map visualizations in Vega-Lite~\cite{satyanarayan2016vega} to chart blocks by extracting corresponding encoding sets from the underlying Vega-Lite specification (see \ref{sec:generalizable} for further discussion of the generalizability of our schema). Similarly, one can write scripts to map Tableau workbooks into our proposed adjacency and interaction graphs, as we will show in the following section.

%% file: sections/04-methods.tex
\section{Methods}\label{sec:tableau_analysis}
The goals of our work are to explore dashboards both \emph{in the wild} (reflective of everyday design practices and analytical needs) and also \emph{at scale} (beyond the feasibility of prior, heavily-manual analysis techniques). In this section, we first provide a brief overview of Tableau's specification format and subsequently summarize the steps we took to analyze a large-scale corpus of Tableau dashboards, including curating dashboards from Tableau Public, converting Tableau dashboards into our intermediate schematic representation, and finally, deriving features for further analysis of dashboard design patterns.

\subsection{Tableau Workbook Specifications}

The processes of winnowing and analyzing the Tableau Public repository is predicated on the specification of workbooks and their dashboards. 
Tableau workbooks (hereon referred to as \textbf{\textit{workbooks}}) are specified using XML and are composed of data sources, worksheets, and dashboard objects\footnote{There is much more content in workbooks, such as metadata about the data source and newly derived data attributes, but these are less pertinent to our analysis and are primarily used to render the workbook in Tableau Desktop or in a web browser.}. \textbf{\textit{Data sources}} store a high-level description of the dataset, including the name and data type of its attributes. 
Individual data visualizations are constructed in \textit{\textbf{worksheets}} by dragging and dropping dataset attributes onto so-called ``shelves'' (i.e., row, column) or to specific encoding channels (i.e., color, size, etc.). A visualization is automatically suggested or user-specified by selecting a mark type. 

Workbooks can contain one or more worksheets and one or multiple worksheets can be arranged into a \textit{\textbf{dashboard}}. The user can also choose to add text, image, or layout elements to the dashboard.
Importantly, a dashboard need not contain a data visualization; a valid Tableau dashboard can consist of just images, text, and/or numerical values.
Moreover, there is no restriction on the number of dashboards within a workbook, i.e., workbooks need not contain any dashboards or may have multiple. Collectively these different data visualization and design elements are encoded as \textit{\textbf{zones}} in a dashboard.
Finally, a user can specify \textit{\textbf{actions}} that add interactivity between dashboard zones, including highlighting, (cross-)filtering, and page navigation.

\subsection{Corpora Curation}
Tableau Public is ``a free platform to explore, create, and publicly share data visualizations online''~\footnote{https://public.tableau.com/app/about}.
Analyzing the full corpora of Tableau Public's is infeasible for several reasons. The primary reason is pragmatic -- scraping Tableau Public for all of its content is against the platform's terms of service. Older workbooks can also be more difficult to access because they exist in formats that are no longer supported. Finally, not all workbooks necessarily contain dashboards with some workbooks containing only a single data visualization (or none at all) and thus are not pertinent to our research focus on dashboard design.

\begin{figure}[t!]
    \centering
    \includegraphics[width=.6\linewidth]{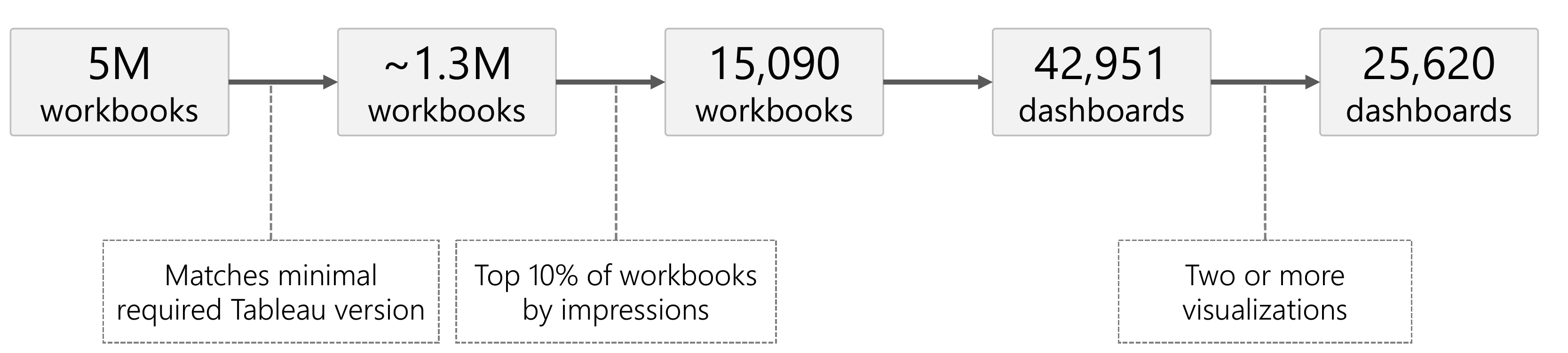}
    \caption{An overview of the dashboard curation process}
    \vspace{-5mm}
    \label{fig:data-pipeline}
\end{figure}

As such, we winnowed the full set of Tableau Public's 5 million analytic workbooks to a set of 25,620 dashboards for analysis; summarized in 
Figure~\ref{fig:data-pipeline}. We leverage the workbooks' XML specification (hereon referred to as specification) to extract the components of dashboards and derive a set workbooks that is suitable for our analysis.
From the total corpus of 5 million workbooks, we limit to workbooks adhering to minimal specification version, a total of 1,342,794 workbooks ($\sim$25\% of all workbooks on Tableau Public); this approach pragmatically resolves issues of older workbooks.
Of these workbooks, only 150,276 (11\%) contained one or more Tableau dashboards.
As a way of improving the potential quality of our corpus, we further filtered workbooks based on \textit{impressions}, a Tableau Public metric tracking the total page views across all dashboards in a workbook.
The impressions distribution across the 150,276 workbooks was left-skewed with a heavy tail with values ranging from just one impression per workbook to over 32 million.
Impressions did not strongly correlate with the publication date.
Given the distribution, we elected to sample the top 10\% of workbooks based on impressions, yielding a set of 15,090 workbooks, with individual workbooks having a minimum of 45 impressions.
These workbooks contained 42,951 Tableau dashboards in total.
As our research focuses on multi-view dashboards, we scoped our exploration to only consider dashboards that have \textit{two or more} data visualizations---applying this criteria results in a final set of 25,620 Tableau dashboards for analysis.
The resulting dashboards were published between 2011 and 2021, but more than 99\% were published after 2014; the 1\% of workbooks in our analysis that were initially published prior to 2014 were recently updated.

\subsection{Deriving Dashboard Graph Representations}

\begin{figure*}[t!]
    \centering
    \includegraphics[width=\textwidth]{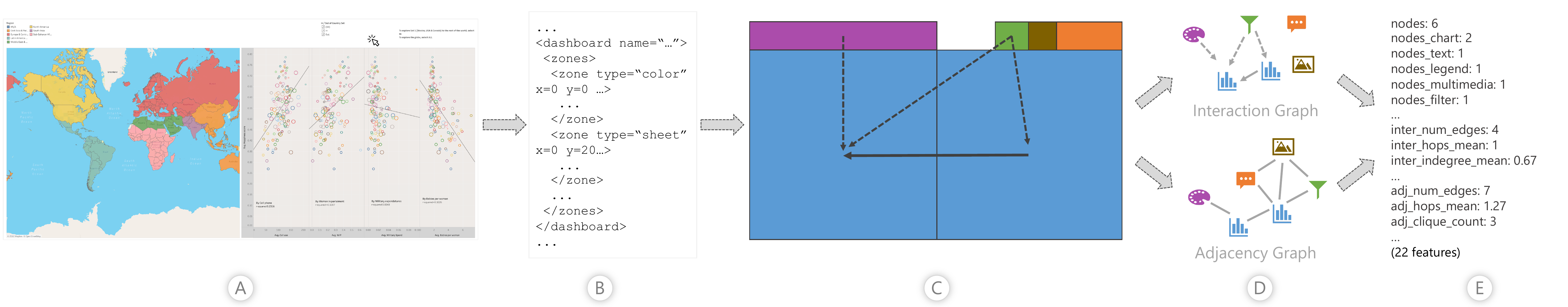}
    \caption{An overview of the feature extraction process. Given a Tableau dashboard (A), we parse the underlying XML specification file (B) to detect the different blocks and connections between blocks (C). We then model two graphs depicting the interactive and spatial configurations of the dashboard (D). From these graphs, we extract 22 features that we use for our analyses (E).}
    \label{fig:feature-extraction}
\end{figure*}

We used the specification of the 25,620 dashboards to construct adjacency and interaction graphs, as defined in~\autoref{sec:formalism}.
An overview of our approach, from parsing workbooks to graph generation and analysis feature extraction, is shown in~\autoref{fig:feature-extraction}.

\subsubsection{Detecting Blocks}
We leverage the notion of ``zone'' objects in Tableau specifications to extract blocks. Parsing these zone objects, we derive five block types: \charts{} containing visualizations, \filters{} containing widgets like dropdown menus and sliders, \legends{} displaying data mappings for graphical encodings like size and color, \textB{} blocks including the dashboard title, caption, or additional commentary, and finally, \multimediaB{} blocks containing images or embedded web pages.
The \chart{} block type links to the original worksheet that describes the data visualization, which we use to extract the visualization type (e.g., bar chart, map, scatterplot, treemap, sankey diagram) from the specified marks (e.g., bar, line, circle) and encodings (e.g., row, column, color).

\subsubsection{Detecting Connections}
Connections can include both interactions between blocks as well as the spatial proximity of blocks in a dashboard.
To detect interaction and adjacency connections, we examine the Tableau dashboard actions and use the spatial positioning of dashboard zones, respectively.

\vspace{.5em}
\noindent\textbf{Detecting Interaction Connections.}
After detecting the dashboard blocks, we extract actions from the specification to define interaction connections between blocks. Each action provides the interaction type as well as the source and target blocks in the dashboard that we use to record connections, as defined in \autoref{fig:data_structure}.
From our workbooks, we are currently able to identify and capture the following forms of interactivity:  {\filter}$\rightarrow${\chart}, {\chart}$\rightarrow${\chart} and {\legend}$\rightarrow${\chart}.

\begin{figure}[b!]
    \centering
    \includegraphics[width=0.45\linewidth]{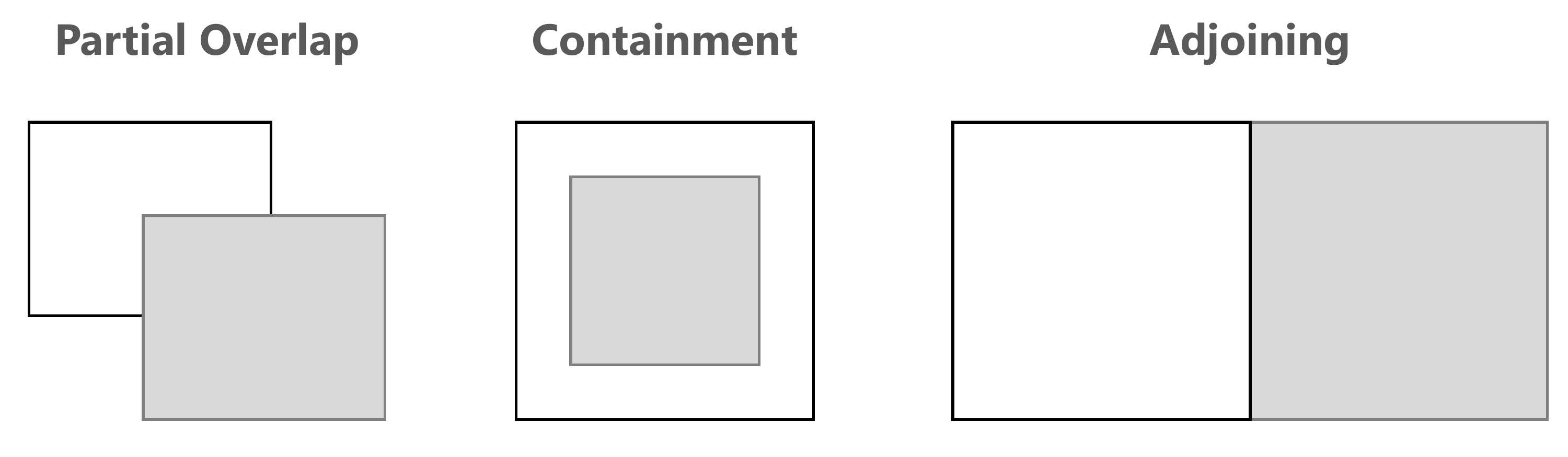}
    \caption{Adjacency configurations between dashboard blocks}
    \label{fig:adjacency-scenarios}
\end{figure}

\vspace{.5em}
\noindent\textbf{Detecting Spatial Connections.}
To construct the adjacency graph, we leverage the coordinate information of blocks to assess spatial proximity. Tableau supports both a grid layout style (the default) and a floating style. However, our method is generalized to other layouts where coordinate information may be extracted. From the grid and floating style layouts, we identified three possible ways that blocks can be adjacent to each other in a dashboard configuration and based on the overlap (partial or total) of their positional coordinates. These adjacency considerations are show in ~\autoref{fig:adjacency-scenarios}.

\vspace{0.5mm}
\noindent\textit{Partial Overlap.} In a floating
layout, two blocks may partially overlap, but neither block is contained entirely within the other.

\vspace{0.5mm}
\noindent\textit{Containment.} In floating dashboard layouts, one block can be contained entirely within another. For example, a {\textTmp} block may be contained entirely with a {\chart} block when it is used to annotate an outlying mark in the data visualization. In this scenario, the coordinate range of one block entirely overlaps with its pair.

\vspace{0.5mm}
\noindent\textit{Adjoining.} 
Primarily in grid layouts
two blocks can share an edge when adjacent to one another (e.g., two \chart{} blocks containing different visualization types could be placed next to one another). Compared to partially overlapping blocks, these adjoining configurations have very limited coordinate overlap, often a few pixels, and requires separate treatment to be accurately detected.

\vspace{.5em}
\noindent{}For a given dashboard, we enumerate all pairwise combinations of blocks and determine if the spatial coordinates of the blocks conform to one of our three adjacency configurations by comparing the rectangular bounding boxes of each block. To allow for flexibility in determining adjacency, we employ a tolerance criterion of approximately 10 pixels that allows two blocks to be positioned a very small distance apart (no shared coordinates) but still be considered adjoining.

\subsubsection{Constructing Adjacency and Interaction Graphs}
We use a similar procedure of enumerating all pairwise combinations of blocks to generate both the interaction and the adjacency graphs. First, each block within a dashboard is considered a node. Second, for each observed connection with the designated connection type, we realize 
an edge between the source and target blocks. While the edges are undirected in the adjacency connection type, they are denoted as directed edges from the source to the target blocks in the interaction graph. 
Finally, we prune the graph edges, eliminating any duplicate edges and self-loops.

\subsection{Extracting Features for Design Pattern Analysis}
\label{sec:clustering-features}

We use the schematic representation of dashboards to explore common dashboard design patterns as \emph{structural similarities} observed across graphs (\textbf{RQ2c}: \textit{``Can we detect and characterize high-level dashboard design patterns?''}).
Here, we describe the features we extract from the computed adjacency and interaction graphs, deferring the clustering analysis and results to the subsequent section.


We generate a total of 22 features for each of the 25,620 dashboards from their corresponding adjacency and interaction graphs. These features contain a combination of features extracted from the nodes, as well as adjacency- and interaction-specific features extracted from the graph links.

\vspace{.5em}
\noindent\textbf{Common Features.}
The number and the types of nodes within both the adjacency and interaction graphs are identical, allowing us to extract a common set of features from both.
These features include the total number of blocks (nodes) in each graph and the presence of specific block types.
We use one-hot encoding to represent the presence (and absence) of block types. 
We standardize total number of blocks by the mean degree, by its mean and unit variance (standard scaling).
We also summarize the total number of edges and the mean degree of nodes within both graphs and apply the standard scaling transform.

\vspace{.5em}
\noindent\textbf{Interaction-Specific Features.} For the interaction graph, we compute the average in-degree and out-degree of the nodes, also applying a standard scaling transformation. We tabulate the presence of the three edge types ({\filter}$\dashrightarrow${\chart}, {\legend}$\dashrightarrow${\chart} and {\chart}$\rightarrow${\chart}) that describe the interactive relationships between two blocks; the presence (and absence) of these edge types are also one-hot encoded.

\vspace{.5em}
\noindent\textbf{Adjacency-Specific Features.} For the adjacency graph, we derive features that add context to the spatial layouts.
For all pairs of blocks in a graph, we compute the average shortest path.
Adjacent blocks have a path length of one.
A path length greater than one indicates multiple ``hops'' are required to reach one block from another and suggests a more complex dashboard layout.
As an additional proxy of layout complexity, we examine adjacency graphs for the presence of one or more \emph{maximal cliques} and, when detected, compute the average size of all cliques.
We apply the standard scaling transformation to the path lengths, number of maximal cliques, and mean clique size.

%% file: sections/05-results.tex
\section{Results}
\label{sec:results}

\begin{figure*}[t!]
    \centering
    \includegraphics[width=\textwidth]{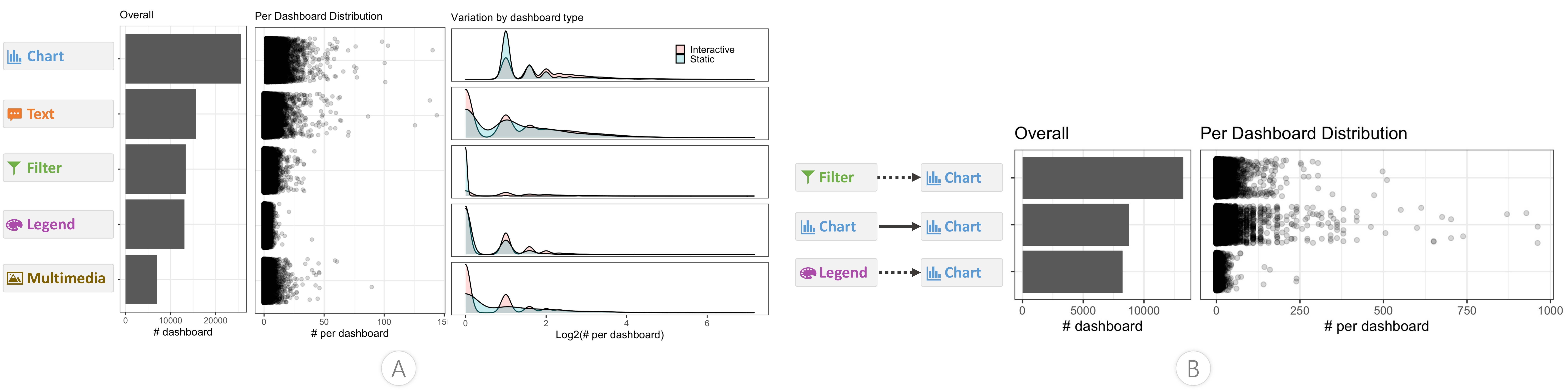}
    \caption{Summary statistics for (A) distribution of block types across dashboards and (B) distribution of interactive connection types across dashboards.}
    \label{fig:quant-summary}
\end{figure*}

\subsection{Visual Composition of Dashboards}
Addressing \textbf{RQ2a}, we examine the visual composition of dashboards focusing on blocks and their spatial relationships.

\subsubsection{What are the visual components of a dashboard?}\label{subsec:dash_composition}
We identified a total of 250,794 blocks across the 25,620 dashboards. The number of blocks per dashboard ranged from 2 to 267 (median: 8, mode: 4). We observed that 121,068 out of 250,794 blocks (49\%) were {\charts}, followed by 53,267 {\textB} blocks (21\%), and subsequently {\filter} (36,472 or 15\%), {\legend} (22,446 or 9\%), and {\multimedia} (17541 or 7\%). We note the importance and centrality of \textB{} and \charts{} as the building blocks of dashboards: together, over 70\% of the blocks in our analysis were one of these two block types.
Figure~\ref{fig:quant-summary}A shows these results in detail.

For the {\chart} blocks, we also examined the distribution of visualization types across dashboards.
We found that bar charts were the most common visualization type, appearing in 15,392 out of 25,620 (60\%) dashboards. The next most frequently used charts were line charts (n=6,524; 25\%), maps (n=6,454; 25\%), and finally, tables (n=6,154; 24\%). Besides other canonical chart types (e.g., scatterplots, pie charts), there were also instances of more bespoke visualizations, such as sankey diagrams and waterfall charts, but they were present in only 116 dashboards ($<$0.5\%).
These findings point to many dashboards in our corpus representing a combination of simple charts (like bar charts) and a smaller set of more sophisticated dashboard designs.

\subsubsection{What are common structural relationships between visual components in a dashboard?}\label{subsec:dash_spatial}

\begin{figure}[ht!]
    \centering
    \includegraphics[width=.5\linewidth]{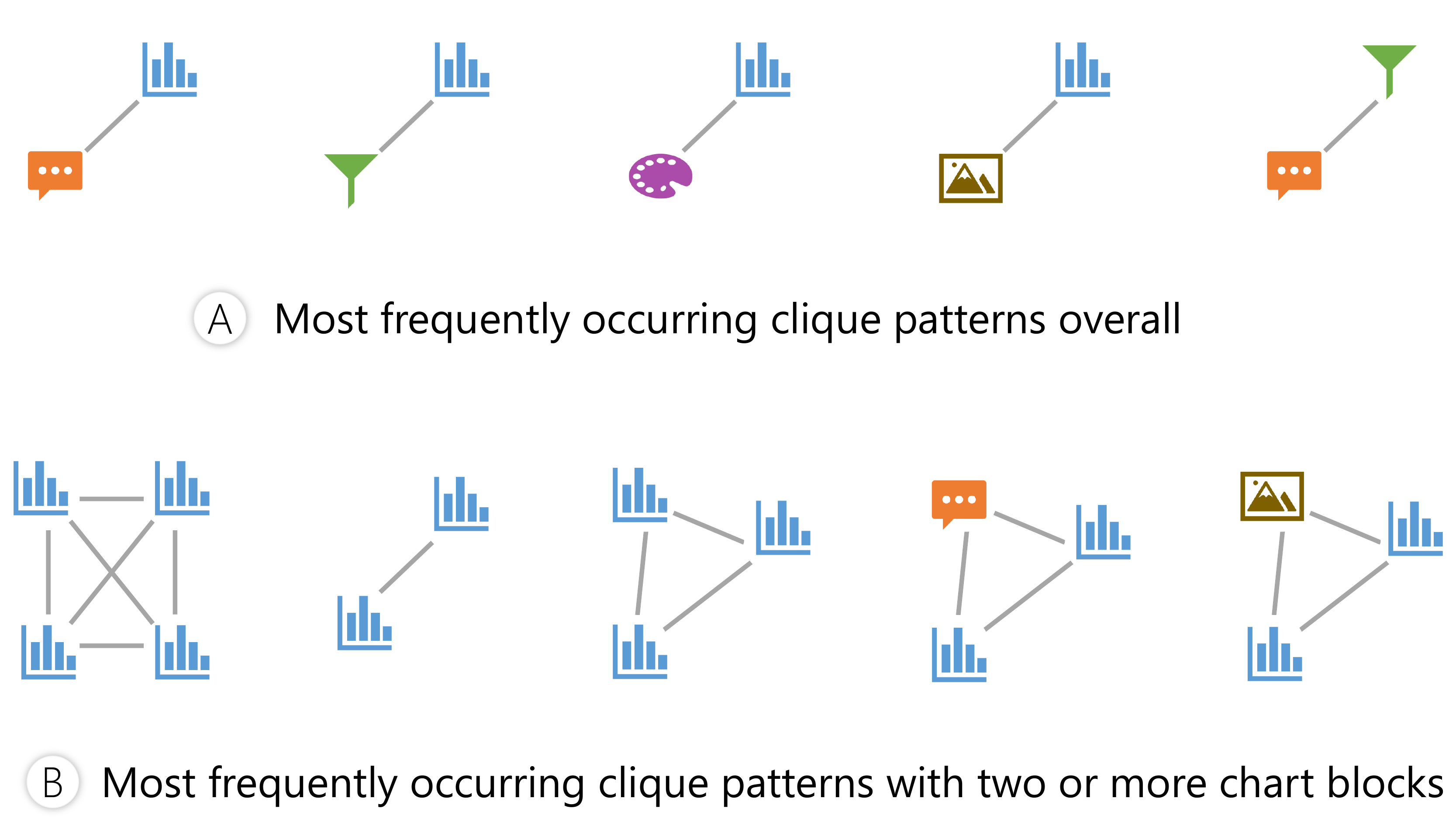}
    \vspace{-2mm}
    \caption{Frequently occurring clique patterns from the adjacency graph analysis. Nodes map to different block types including {\chart}, {\textTmp}, {\filter}, {\legend}, and {\multimedia}.}
    \label{fig:clique-summary}
\end{figure}

In addition to the analyses of the properties of individual blocks, we also explored spatial relationships between blocks. 
Specifically, for the 25,620 dashboards, we analyzed the adjacency graphs to identify potential design patterns around block layouts.
The graphs had 2-267 nodes (median: 8, mode: 4) and 0-4926 edges (median: 11, mode: 3). For each graph, we extracted the list of all maximal cliques, capturing the block type for each node in a clique (e.g., \{\chart$-$\chart$-$\chart\}, \{\chart$-$\filter\}, \{\multimedia\}).
Aggregating these maximal clique patterns across all graphs, we found a total of 1,430 unique block patterns. The smallest-sized cliques contained just 2 blocks and the largest contained 60, and the median clique size was 9. The most frequently occurring clique patterns contained just two blocks, typically including a {\chart} and one of the other block types; a summary of the most frequently occurring patterns is in Figure~\ref{fig:clique-summary}A.

Focusing on {\chart} blocks specifically, we also examined the creation of juxtaposed views.
Of all 1,430 unique clique patterns, more than half (n=747) contained a spatial arrangement of two or more {\charts}; we list the five most frequently occurring clique patterns in Figure~\ref{fig:clique-summary}B.
The most dominant patterns are cliques containing only {\chart} block types, varying in size from 2 to 4.
When two {\chart} blocks occur with other blocks, it was more common for those other blocks to be a {\textTmp} or {\multimedia} block.

Collectively, the clique analysis shows that \textit{like} is often juxtaposed with \textit{like} in dashboards: common dashboard elements are visually grouped together.
As with the aforementioned block composition analysis, the results emphasize simpler dashboard designs, but, reiterate that authors do experiment with more complex designs that mix block types within close spatial proximity, such as having {\textTmp} or {\multimedia} blocks that are connected to multiple {\charts}.

%% file: sections/05b-interaction-analysis.tex
\subsection{Interactivity in Dashboards}
Complementing our analysis of the visual components of dashboards (\textbf{RQ2a}), we also analyzed the connections between blocks to understand what types of interactions dashboards commonly support (\textbf{RQ2b}).
Note that we refer to a dashboard as \emph{interactive} if clicking on one block updates another block (e.g., by filtering, highlighting, or changing visualized data fields), as opposed to other forms of interactivity that only involve individual blocks (e.g., hovering over a point in a single chart to generate a tooltip).

\subsubsection{Is interactivity common in dashboards?}\label{subsec:dash_interaction}
We found that 19,304 of the 25,620 dashboards in our corpus (75\%) were interactive and that their design patterns varied considerably. In particular, the number of interaction connections between blocks ranged from 1 (e.g., a filter acting as a control for a single chart) to 992 (median: 6, mode: 2) interaction connections in a dashboard. Recall that a single block can be the source or target of multiple interactions.
The maximum possible interaction connections in a Tableau dashboard is bound by \begin{math}(\charts-1 + \legend + \filters) * \charts\end{math}.
On average, 58\% of the possible interactions between blocks were applied (median: 50\%, mode: 100\%), suggesting that when authors add interactions, they tend to make a considerable portion of the dashboard interactive.

\subsubsection{How does interaction commonly manifest?}
Of the three Tableau blocks that support interaction connections (\chart{}, \filter{}, and \legend{}), {\filter}$\dashrightarrow${\chart} was most common with 13,228 out of the 19,304 interactive dashboards (69\%) supporting this style of interaction, followed by {\chart}$\rightarrow${\chart} (46\% of dashboards) and {\legend}$\dashrightarrow${\chart} (43\% of dashboards). These results are summarized in Figure~\ref{fig:quant-summary}B and suggest that there are multiple strategies for interaction design or entry points for users of an interactive dashboard.

\begin{figure}[t!]
    \centering
    \includegraphics[width=.8\linewidth]{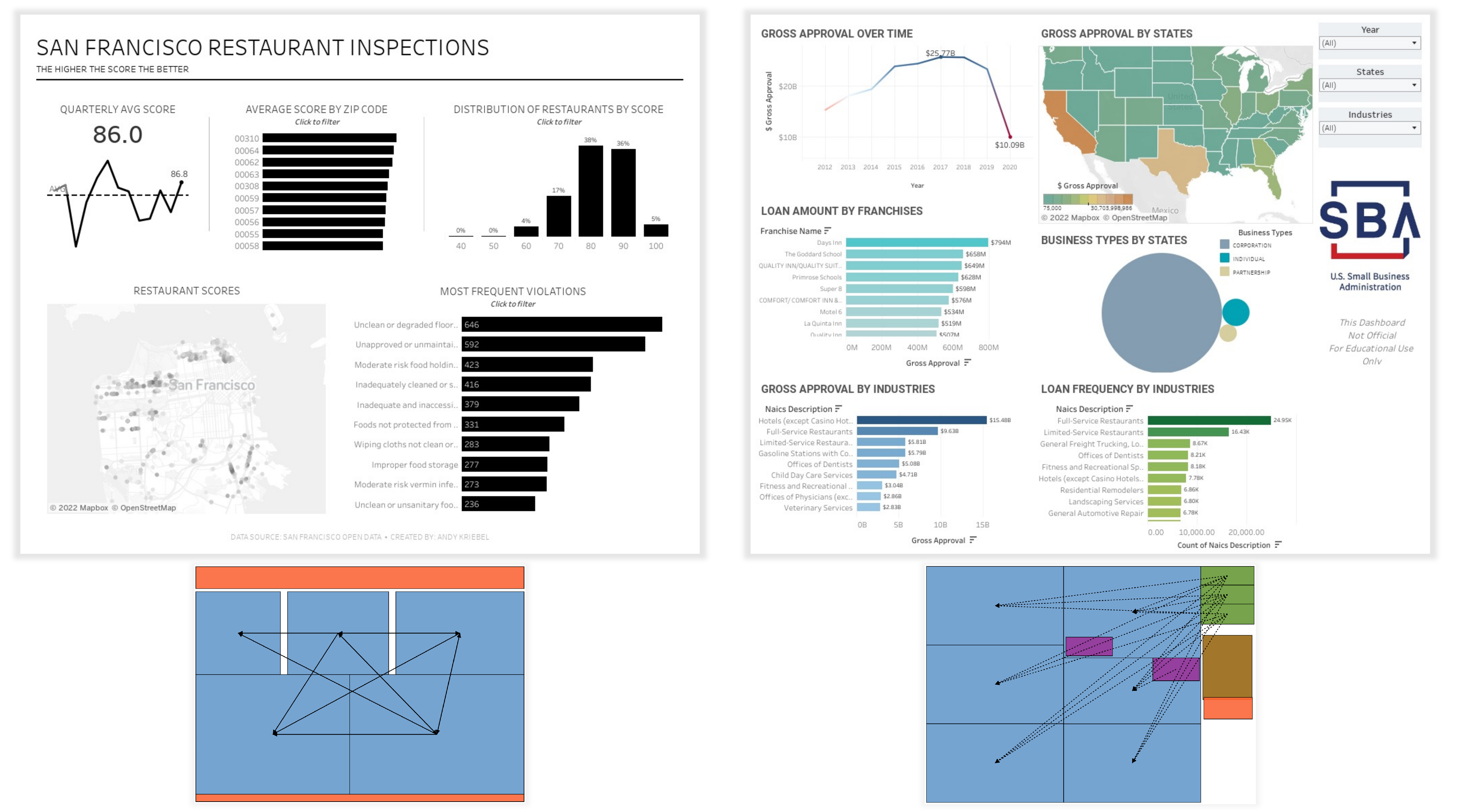}
    \caption{Stereotypical examples of dashboards (along with their underlying blocks and connections) illustrating two prevalent interaction patterns we observed in our corpus: (Left) a highly interlinked group of \charts{} and (Right) a control panel-style dashboard with \filters{} and \legends{} driving the \charts{}.}
    \label{fig:interaction-pattern-examples}
\end{figure}

Collecting both spatial adjacency and interactivity in graph structures sharing common nodes allows us to assess the relationship between these two factors.
While one might assume that the blocks that \textit{control} a particular portion of the dashboard would be spatially next to each other, we found that this was not always the case.
In fact, out of a total of 343,929 interactions, only 105,317 (30\%) were in cases where blocks were adjacent.
Out of these 105,317 adjacent + interactive connections, 58275 (55\%) were {\chart}$\rightarrow${\chart} interactions, 37186 (35\%) were \filter{}$\dashrightarrow$\chart{}, and the remaining 9856 (10\%) were \legend{}$\dashrightarrow$\chart{} interactions.
This distribution suggests two broad genres or patterns of interaction design: one were cliques or tightly connected subgroups of \charts{} mutually interact (\autoref{fig:interaction-pattern-examples}-left), and another ``light switch'' style dashboard where a control panel with \filters{} and \legends{} interacts with many if not all charts on a dashboard (\autoref{fig:interaction-pattern-examples}-right).

\begin{figure}[t!]
    \centering
    \includegraphics[width=.5\linewidth]{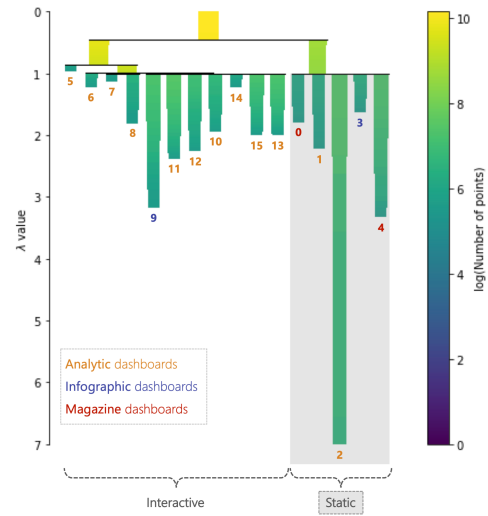}
    \caption{Dendrogram generated by the HDBSCAN algorithm summarizing cluster hierarchies. Cluster IDs are colored to match the design patterns they were mapped to. The distinct split between clusters 0-4 on the right and clusters 5-15 on the left illustrates that the algorithm picked up on the presence (or absence) of interactions as a salient feature for clustering.
    }
    \label{fig:hdbscan-dendrogram}
\end{figure}

\subsection{Characterizing Clusters of Dashboard Design Patterns}
\label{sec:cluster-analysis}

In addition to summarizing the visual composition and interactivity in our dashboard corpus, we also explored if the proposed adjacency and interaction graphs can be used to detect high-level dashboard design patterns (\textbf{RQ2c}).
Specifically, we used the features derived from the graph representations (Section~\ref{sec:clustering-features}) to perform an unsupervised cluster analysis using the hierarchical density-based clustering (HDBSCAN)~\cite{Campello:hdbscan:2013} algorithm.
We chose HBDSCAN over other clustering algorithms, such as k-means or Mixture Models, since it allows for some points to remain unclustered---essentially to be viewed as ``noise.''
This behavior enables identifying more consistent clusters of design patterns as the algorithm does not force dashboards into clusters, which can be important when programmatically analyzing designs at scale.
For instance, if a dashboard's design is distinctive relative to all other dashboards or alternatively if a dashboard could justifiably belong to multiple clusters, the algorithm can flag it as noise and preserve coherence within the other clusters.
Conversely, approaches like k-means would force all dashboards into clusters, adding noise to the emergent design patterns.
We apply a conservative criterion for clustering, selecting a minimal cluster size of 250 (roughly 1\% of the overall corpus), a number we determined through sensitivity analysis\footnote{Additional details are provided in the supplemental material.}.

\begin{figure*}[ht!]
    \centering
    \includegraphics[width=.9\textwidth]{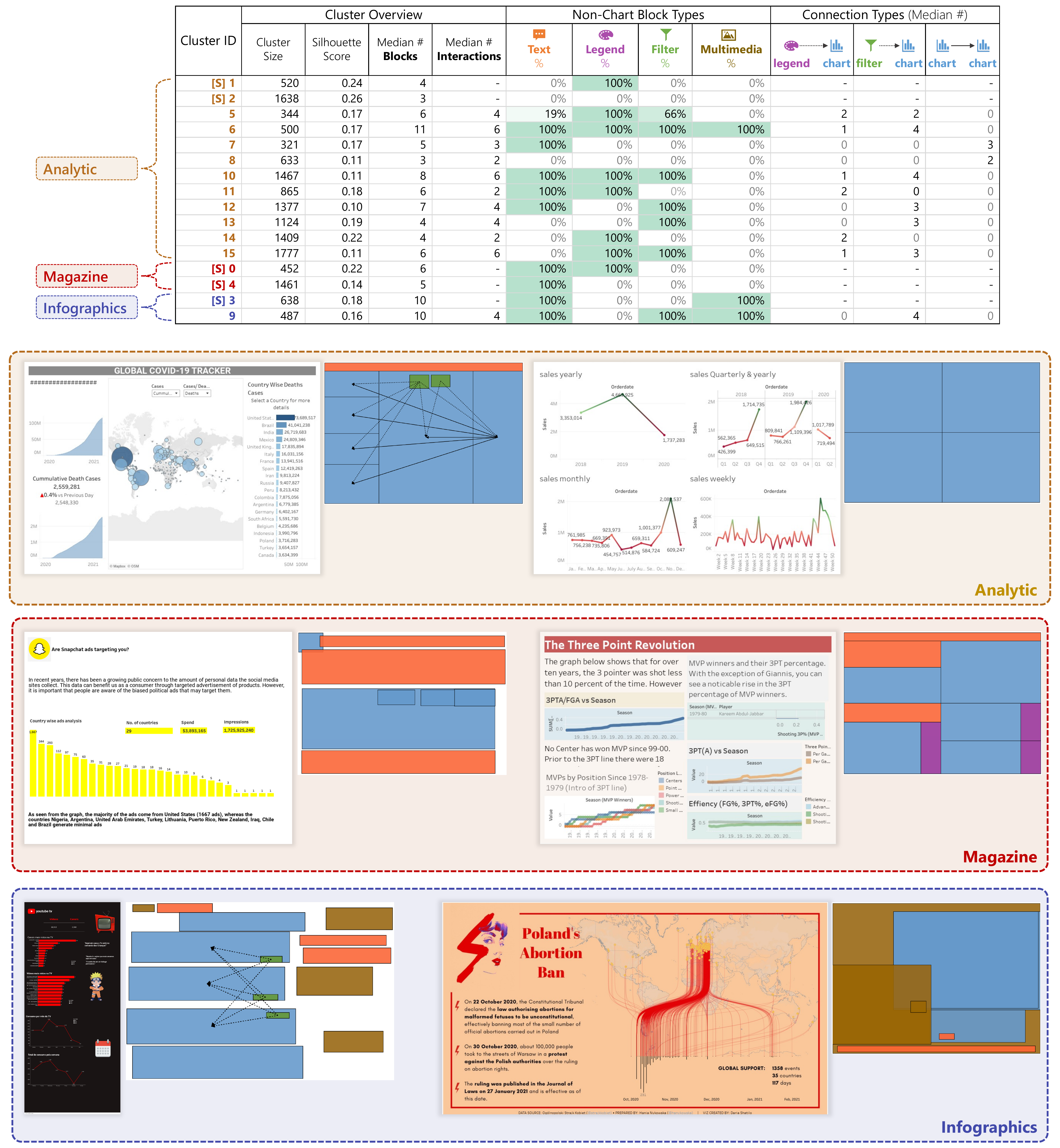}
    \caption{Summary statistics of the 16 clusters detected using HDBSCAN (Top).
    Cluster IDs prefixed with `[S]' contain only static dashboards.
    Groups of clusters mapped to three high-level dashboard patterns namely including \textit{Analytic dashboards}, \textit{Magazine dashboards}, and \textit{Infographic dashboards}.
    (Bottom) Two sample dashboards corresponding to each pattern along with their underlying block and connection configuration.}
    \label{fig:clustering-results}
\end{figure*}

The algorithm detected 16 clusters covering 15,013 out of the 25,620 (59\%) dashboards in our corpus, while the remainder was flagged as noise.
Figure~\ref{fig:clustering-results}-top summarizes these 16 clusters using summary statistics about the cluster size and silhouette scores, as well as the blocks and connections within each cluster.
These summary statistics help identify the general design of dashboards within each cluster. For instance, we can note that clusters such as 6, 9, and 10 contain a mix of block types, whereas clusters 2 and 8 contain only \charts.

To go beyond the design properties of individual clusters and identify higher-level dashboard design patterns in the corpus, we first analyzed the relationships between clusters by inspecting the cluster hierarchy dendrogram produced by HDBSCAN (Figure~\ref{fig:hdbscan-dendrogram}). 
The dendrogram highlighted a clear distinction between static and interactive dashboard clusters and also pointed us to potentially related clusters (based on their ordering in the dendrogram).

Complementing the algorithmic approach of the dendrogram, we then manually examined the clusters. Specifically, for each cluster, we looked at a set of representative dashboards and inspected their designs.
Based on the visual properties of the dashboards, we mapped each cluster to a recently proposed set of dashboard design patterns by Bach et al.~\cite{bach2022dashboard}.
We detected three classes of dashboards design patterns including \textit{analytic dashboards}, \textit{magazine dashboards}, and \textit{infographic dashboards}.
Analytic dashboards were the most common type and typically include a set of \charts{} covering different aspects of the underlying data as their primary content. Magazine dashboards typically include multiple \textB{} blocks that complement the \charts{} and provide additional commentary about the data and key takeaways. Finally, infographic dashboards generally include a rich mix of block types including at least one \multimedia{} and \textB{} block in addition to \charts{}.

Note that these design patterns are neither mutually exclusive (i.e., a dashboard can be considered both an analytic and infographic one) nor is our goal to provide an exhaustive list of dashboard patterns for Tableau dashboards. Rather, the purpose of this analysis is to 1) showcase that the proposed schematic representation of dashboards affords the programmatic generation of a logical set of dashboard design clusters, and 2) illustrate that the programmatic approach effectively complements existing predominantly manual efforts of analyzing dashboards, making it possible to characterize dashboard design patterns at scale.


%% file: sections/06-applications.tex
\section{Applications}


The schematic graph representations and the curated corpus of 26,620 dashboards are central contributions of this research.
To lay the groundwork and help foster ideas for future research and development, we describe some exemplary use cases and applications of our representation and dataset.

\subsection{Designing Dashboard Recommendation Tools}
There is a growing interest in visualization recommendations in the context of multi-view visualizations and dashboards (e.g.,~\cite{bach2022dashboard,chen_composition_2021,wu_learning_2021,ma_ladv_2021,kristiansen_semantic_2022}).
For instance, Chen et al.~\cite{chen_composition_2021} present a tool that suggests multi-view visualization layouts by matching a given set of views to a manually annotated dataset of 360 multi-view visualizations.
Our dataset and graph representations can expand this idea at scale and incorporate higher support for interactivity.
Specifically, treating dashboard search as a graph matching task, authoring systems can compare the schematic graphs for an input dashboard to graphs in the dashboard corpus and recommend examples of similarly designed dashboards.
Because we consider both adjacency and interaction graphs, such systems can go beyond prior work that focuses largely on layouts and instead recommend dashboards based on both layout and/or interactive configurations between blocks.

\subsection{Developing Dashboard Linters}



Poor choices in dashboard design or layout can produce designs that are confusing or even misleading.
In particular, Qu \& Hullman~\cite{qu2017keeping} suggest that keeping multiple views ``consistent'' is important for the legibility of dashboards.
In our corpus, we observed occasional mistakes or violations in dashboard design that were visible through inspection of the \emph{graph structure alone}.
For instance, interactive {\filters{}} that were centrally placed and only applied to some (but not all) of the {\charts{}} in a dashboard, a commonly placed \legend{} although two charts used a different color mapping, etc.
This suggests the ability for our graph formalisms to be used to automatically ``lint'' or otherwise ``audit''~\cite{chen2022vizlinter,mcnutt_surfacing_2020} dashboards and surface potential issues during dashboard authoring.

\subsection{Inferring Dashboard Intent and Templates}

Prior work has shown that dashboards are generated for a variety of purposes and that intents play an integral role in dashboard design~\cite{sarikaya_what_2019,pandey2022medley}.
However, the process of inferring dashboard intent in their work has been largely qualitative and performed at a small scale.
For instance, Pandey et al.~\cite{pandey2022medley} derive dashboard intents such as ``change analysis'' and ``category analysis'' by manually inspecting the views, filtering widgets, and textual content of 200 dashboards.
Our schematic representation and dataset present an opportunity to investigate this idea at scale and explore how a combination of information from blocks (\charts{}, \textB{}, and \filters{}) and their connections can be used to programmatically infer dashboard intent.
Furthermore, the ability to infer intents and map them to both the positional properties of individual blocks (x, y, width, height) and the connections between blocks presents an opportunity to compile intent-based design templates that incorporate smart defaults for both layout and interaction.

\subsection{Understanding Tool-specific Authoring Practices}

Adopting the schematic representation and performing the described analyses on dashboards authored using a particular tool, can help reflect on the use of specific features and understand potential authoring pain points.
For instance, Tableau allows composing dashboards using a default grid layout while also providing the option for specifying customized layouts.
Merely inspecting the block diagrams in \autoref{fig:clustering-results} confirms that authors use both layout options and suggest that the grid layout is more commonly used in analytic dashboards, whereas customized layouts are more common for infographic- and magazine-style dashboards.
Besides layouts, the blocks of the second infographic dashboard example in Figure~\ref{fig:clustering-results} highlight an interesting case where the textual content on the bottom-left corner is actually specified using a \multimedia{} (image) block.
Upon closer inspection, we can notice that the textual content uses a specific font and formatting and has iconified bullets, which are challenging to create directly within Tableau.
Leveraging block types and their distributions can similarly help identify other authoring ``hacks.''
Identifying and reflecting on such authoring practices can ultimately help better understand user needs and inform features that make it easier for authors to fluidly create desired dashboard artifacts without having to switch tools.

%% file: sections/07-discussion.tex
\section{Discussion}
\label{sec:discussion}

\subsection{From a ``Petting Zoo'' to a ``Census'' of Dashboards}



Prior research that examines dashboards primarily use qualitative approaches to annotate dashboard features and derive design genres or patterns. However, applying the outputs (e.g., guidelines or frameworks) from these prior works at scale remains a challenge.
Without an effective representation of the dashboard, it is not possible to broadly identify these genres and patterns ``in the wild''.
Moreover, as only a limited number of dashboards can be reasonably analyzed manually, the findings from these qualitative approaches can inadvertently prioritize the best and most polished practices that rise to the top of web searches or pique personal interest. As with Heer et al.'s~\cite{heer2010zoo} ``visualization zoo'', these hand-picked and hand-analyzed corpora may not reflect common practice or dashboards of everyday use: ``After all, you don't go to the zoo to see chihuahuas and raccoons; you go to admire the majestic polar bear, the graceful zebra, and the terrifying Sumatran tiger.''

Our scalable approach allows us to move beyond the zoo and towards a census. In our census, we show that dashboard quality varies considerably, with many being far simpler than prior research accounts for. While we cannot identify a singular mechanism behind this variability or this bias toward simplicity, potential rationales suggest missing areas in our current thinking and understanding of dashboard design and use. For instance, existing authoring tools may not offer sufficient support for the majority of dashboard authors to create richer dashboards. New authoring paradigms or tools could address this mismatch. Or alternatively, the analytical needs and data literacy of dashboard audiences may be fully met by simple and static collections of one or two simple charts: if so, then research that focuses on more ``exotic'' forms of visual presentation and interactivity may fail to meet users where they are and assist them with their everyday analytical goals.


Our approach is not without its limitations, but when its results are taken together with those of prior research, we gain a more complete and diverse perspective on dashboard authoring practices. Moreover, this diversity is important, as there is growing interest in applying machine learning and other automation techniques to dashboard creation. Understanding and describing the full spectrum of dashboards becomes critical to ensure authoring support and that recommendations are relevant. 

\subsection{Implications for Dashboard Authoring System Design}

\subsubsection{Improved Support for Non-Chart Blocks}
Dashboards have conventionally been considered as visual analytic artifacts composed predominantly of multiple coordinated views~\cite{few2006information}.
However, our analysis shows that, in practice, non-\chart{} blocks such as \textB{} and \multimedia{} (e.g., images) play an integral role in dashboard design. An important consideration for future dashboard tools is to provide ample support for authoring and incorporating such content in flexible ways. The prevalence of \textB{}, in particular (\autoref{fig:quant-summary}A), also hints at potential synergies with recent research on interactively linking text and charts~\cite{latif2021kori,badam2018elastic,zhi2019linking}, and presents an opportunity to further explore this relationship in the context of dashboards.


\subsubsection{Prioritize support for customization of simple charts over authoring bespoke charts.}

As stated in Section~\ref{subsec:dash_composition}, our analysis showed that basic visualizations, including bar charts, maps, line charts, tables, pie charts, etc. are more prevalent in dashboards with only a minor subset of dashboards containing bespoke charts.
However, while inspecting examples during the cluster analysis, we observed that the absence of bespoke charts did not dampen the ``richness'' of dashboard designs and that authors either heavily format basic charts or combine basic charts such as maps and pie charts in innovative ways to create visually compelling designs.
Unfortunately, creating such highly stylized and custom designs with tools like Tableau can require substantial expertise, forcing new or novice authors, in particular, to resort to default visuals or integrate artifacts across design and visualization tools.
As the user base for dashboard design tools broadens, it is more important to provide more expressive and flexible authoring interfaces for formatting (including using images and icons as marks~\cite{haroz2015isotype}) and integrating basic charts than to focus on allowing dashboard authors to create and incorporate more bespoke visualizations.

\subsubsection{Need for Better Onboarding Techniques to Support Interaction.}

75\% of the dashboards in our corpus were interactive and supported using at least one block to visually update one or more other blocks.
We also found that interactions are configured between different pairs of blocks (\filter{}$\dashrightarrow$\chart{}, \chart{}$\rightarrow$\chart{}, \legend{}$\dashrightarrow$\chart{}) and are not always consistently used in a dashboard (e.g., \filters{} may only update a subset of \charts{}).
While this nuanced configuration of interactions suggests that current tools provide a rich set of features to author interactivity, it raises questions about the discoverability and usability of such dashboards from a viewer standpoint. One important direction for future systems is to incorporate built-in strategies to improve the dashboard onboarding process~\cite{dhanoa_process_2021} and orient viewers to a dashboard and its use (e.g., through overlaid walkthroughs, or using assistive tooltips and guiding text). Notably, the proposed graph representations can be valuable for designing such features too as the underlying links can be used to identify the flow of actions between blocks.

\subsection{Generalizability beyond Tableau and Considering Alternative Dashboard Definitions}
\label{sec:generalizable}

All results presented in this paper are based on the corpus of 25,620 Tableau Public dashboards, and our analyses are influenced by Tableau's specification and dashboard authoring features.

\begin{figure}[t!]
    \centering
    \includegraphics[width=.4\linewidth]{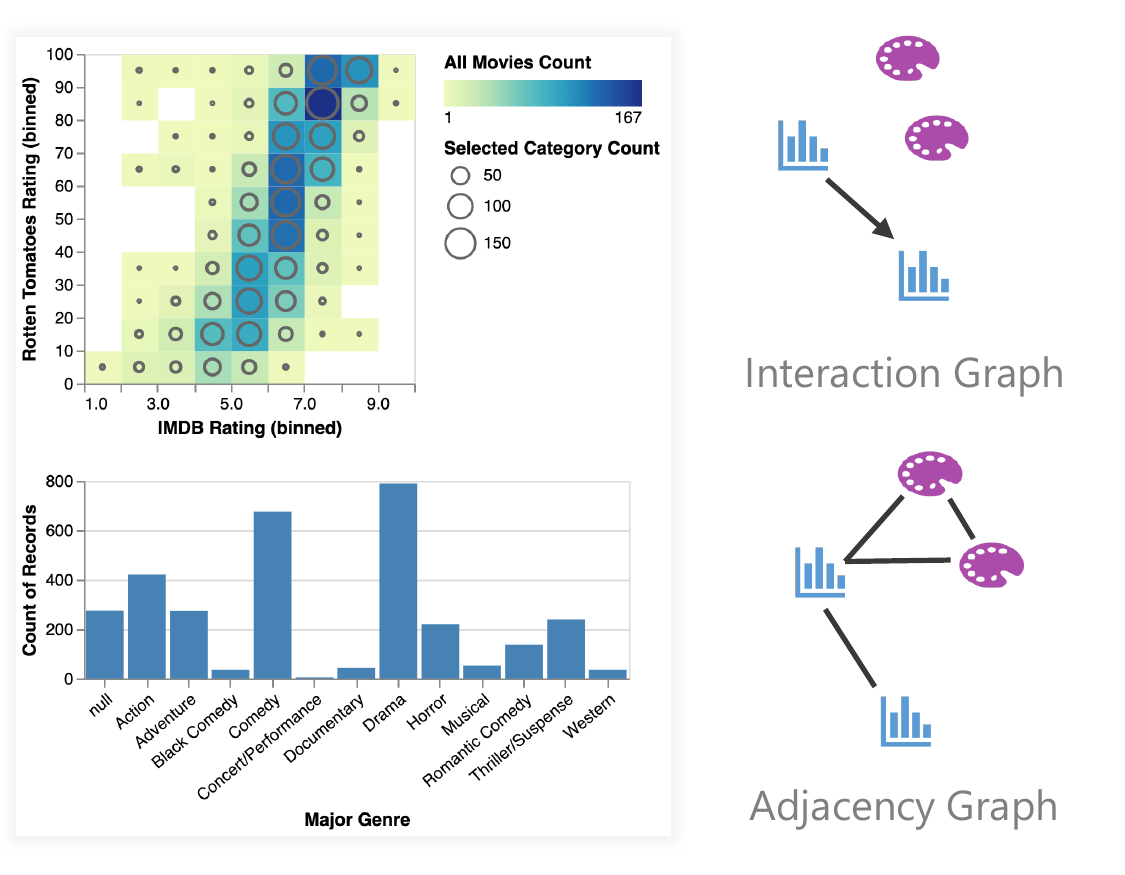}
    \caption{An example of interaction and adjacency graphs extracted from a dashboard specified using Vega-Lite~\cite{vldashboard}.
    Such graphs can be added to the presented corpus to compare or collectively analyze dashboards created using Vega-Lite and Tableau.
    }
    \label{fig:vega-lite-eg}
\end{figure}

However, the proposed idea of blocks and connections, and the corresponding graph representations are based solely on the constituent elements of dashboards and is thus agnostic to \emph{how} these elements were created. This representation provides sufficient breadth to capture a wide range of design elements, from any type of visualization and text element to even non-visualization elements such as web pages. 
As a result, our schematic representation can capture the core elements of \emph{any existing dashboard}, whether this dashboard is specified manually using graphic design tools, automatically through recommendation systems, or programmatically using visualization languages (e.g., \autoref{fig:vega-lite-eg}).
For dashboards that exist in other media, such as print, we can use alternative techniques such as image segmentation~\cite{huang2007,poco2017,poco2018,savva2011} to identify blocks and connections.
Ultimately, the value of this representation lies in how it provides a consistent \emph{intermediary representation} for dashboards that transcends individual tools.
With a single representation, we can capture and analyze virtually any existing dashboard design, enabling our community to develop future corpora not as isolated islands but as inter-operable and comparable datasets amenable to rich meta-analysis.

We also limited our search to dashboards that had \textit{two or more} visualizations (\autoref{fig:data-pipeline}).
This resulted in us having ample examples of dashboards that fit prior definitions of the term as commonly used in visualization research (e.g., ``\textit{a visual data
representation structured as a tiled layout of simple charts and/or large numbers}''~\cite{sarikaya_what_2019}).
As this criteria for defining a dashboard is adjusted to include additional examples---for instance, those having a single visualization with multiple filters, so will the results of the design analysis.
To reiterate, our formalism and corresponding analysis approach would still remain applicable and could be reused to understand the design of an updated dashboard corpus with a more stringent or lax scoping.

%% file: sections/08-conclusion.tex
\section{Conclusion}


We present a corpus of 25,620 dashboards from Tableau Public and analyze this corpus based on
patterns of visual design, layout, and interactivity.
Rather than considering a manually-selected corpus of hand-picked examples, our work
captures a wide set of dashboards created by a variety of authors and for different purposes. Besides highlighting high-level dashboard design patterns, our analysis also brings to focus important areas of dashboard design that have often been overlooked, such as the importance of non-visualization elements, the importance and diversity of interactions, and the challenges that authors face when building interactive dashboards for their intended audience.
Finally, we make our dashboard corpus publicly available\footnote{\url{https://osf.io/r5cfk/?view_only=0cca76d27dac402496646c20359c3da6}}
and discuss its potential application toward exploring new and exciting research directions in both supporting and understanding the dashboard authoring process.